\documentclass[12pt,preprint]{aastex}
\usepackage{amsmath}

\shorttitle{Structure of B335}
\shortauthors{Harvey et al.}

\begin{document}

\slugcomment{Accepted by the ApJ}

\title{Inner Structure of Protostellar Collapse Candidate B335\\
Derived from Millimeter-Wave Interferometry\footnote{Based on
observations carried out with the IRAM Plateau de Bure Interferometer.
IRAM is supported by INSU/CNRS (France), MPG (Germany) and IGN (Spain).}
}

\author{Daniel W.A.\ Harvey, David J.\ Wilner, Philip C.\ Myers}

\email{dharvey, dwilner, pmyers@cfa.harvard.edu}

\affil{Harvard-Smithsonian Center for Astrophysics, 60 Garden Street,
Cambridge, MA 02138}

\author{Mario Tafalla}
\email{tafalla@oan.es}
\affil{Observatorio Astronomico Nacional, Apartado 1143,
  E-28800 Alcala de Henares, Spain}
\and
\author{Diego Mardones}
\email{mardones@das.uchile.cl}
\affil{Departamento de Astronomia, Universidad de Chile,
   Casilla 36-D, Santiago, Chile}

\begin{abstract}
We present a study of the density structure of the protostellar collapse  
candidate B335 using continuum observations from the IRAM Plateau de Bure 
Interferometer made at wavelengths of 1.2~mm and 3.0~mm. 
We analyze these data, which probe spatial scales from 5000~AU to 500~AU, 
directly in the visibility domain by comparison to synthetic observations
constructed from models that assume different physical conditions. 
This approach allows for much more stringent constraints to be 
derived from the data than from analysis of images.
A single radial power law in density provides a good description of 
the data, with best fit power law density index $p=1.65 \pm 0.05$.
Through simulations, we quantify the sensitivity of this result to various 
model uncertainties, including assumptions of temperature distribution, 
outer boundary, dust opacity spectral index, and an unresolved central 
component. The largest uncertainty comes from the unknown presence
of a centralized point source. The maximal point source with 1.2~mm flux of 
$F=12 \pm 7$~mJy reduces the power law density index to $p=1.47 \pm 0.07$.
The remaining sources of systematic uncertainty, of which the most important
is the radial dependence of the temperature distribution, likely contribute
a total uncertainty at the level of $\delta p \lesssim 0.2$.
Taking account the uncertainties, we find strong evidence that the power  
law index of the density distribution within 5000~AU is 
significantly less than the value at larger radii, close to 2.0 from 
previous studies of dust emission and extinction. 
Images made from the data show clear departures from spherical symmetry,
with the globule being slightly extended perpendicular to the outflow axis.
The inclusion of a crude model of the outflow as a hollowed bipolar 
cone of constant opening angle improves the fit and leaves the resulting  
density power law index unchanged. 
These results conform well to the generic paradigm of isolated, 
low-mass star formation which predicts a power law density index close to 
$p=1.5$ for an inner region of gravitational free fall onto the protostar. 
However, the standard inside-out collapse model does not fit the data as  
successfully as a simple $p=1.5$ power law because of the relative 
shallowness of the predicted density profile just within the infall radius. 
\end{abstract}


\section{Introduction}

The dense core in the B335 dark globule is generally recognized as 
the best protostellar collapse candidate. This dense core is nearby 
(250~pc, Tomita, Saito \& Ohtani 1979), isolated, and nearly spherical.
It contains a deeply embedded low luminosity young stellar object 
($3$~L${}_{\odot}$) discovered at far-infrared wavelengths by 
Keene et al.\ (1983) and detected by IRAS only at $\lambda \geq 60$~$\mu$m.
Detailed radiative transfer models based on the theory of 
inside-out collapse (Shu 1977) provide very good fits to 
spectral line profiles of the dense gas tracers CS and H$_2$CO observed 
at $10''$--$30''$ resolution (Zhou et al.\ 1993; Choi et al.\ 1995).

Recent studies have raised serious doubts about the inside-out collapse 
interpretation of the molecular line profiles. Observations of the CS(5-4) 
line at higher angular resolution, about 500~AU, show that the high
velocity emission arises from the inner part of a bipolar outflow, not 
from the gravitational acceleration of dense gas close to the
protostar (Wilner et al.\ 2000). In addition, studies of starless cores 
show that some molecular species, in particular CS, are frozen out onto 
grains at densities characteristic of the inner regions of B335 
(Caselli et al.\ 2002, Tafalla et al.\ 2002). 
If protostellar heating does not promptly desorb CS molecules  
from icy grain mantles, then this species likely makes a poor probe 
of the gas kinematics in the presumed infall zone. 

Since the density field is strongly coupled to the velocity field,
observations of dust emission and absorption provide an alternative
means to infer the dynamical state of the dense core.
Harvey et al.\ (2001) describe deep near-infrared extinction measurements
toward B335 that show a density profile consistent with the
inside-out collapse theory, in particular an $r^{-2}$ density profile for 
the outer envelope and an inner turnover towards free-fall. The best fit
model has an infall radius of 6500~AU ($26''$), similar to that 
derived from molecular line studies (Zhou et al.\ 1993; Choi et al.\ 1995).
The extinction data also show the 
dramatic effect of the bipolar outflow driven by the protostar.
However, extinction measurements could not be made interior to $3500$~AU  
($14''$), where no background stars are visible, even with NICMOS on HST.
As a result, the extinction data could not distinguish the inside-out 
collapse model from, e.g. a highly unstable Bonnor-Ebert sphere, 
whose density profiles depart significantly from each other only at 
radii $\lesssim 2500$~AU. 

Observations of long wavelength dust emission provide another method 
for probing density structure. This technique is nearly as direct as 
observations of dust extinction. The intensity of the emission provides 
an integral along the line-of-sight of the product of the density, 
temperature, and opacity of the dust. By modeling the dust temperature and 
specific mass opacity, the observed intensities may be used to constrain 
the density distribution. The analysis of dust emission complements 
extinction work because the dust emission method becomes most effective 
in the high column density inner regions of dense cores where dust
extinction becomes large and difficult to penetrate.

Dust emission has been used very successfully to measure density structure
of dense cores using bolometer cameras at millimeter and submillimeter
wavelengths on large single dish telescopes, including the IRAM 30~meter  
(Ward-Thompson, Motte \& Andre 1999, Motte \& Andre 2001) and the 
JCMT (Ward-Thompson et al.\ 1994; Visser, Richer \& Chandler 2001,
Shirley et al.\ 2000). Unfortunately, the central regions of the nearest
protostellar cores are generally comparable in size to the beamwidths 
of these telescopes and are poorly resolved. The density structure at 
smaller scales can only be probed with interferometers. With 
interferometer observations, the density structure may be strongly 
constrained through analysis of dust emission directly in the visibility 
domain. So far, detailed modeling of millimeter continuum visibility data 
in this context has been rarely applied (Keene \& Masson 1990, 
Hogerheijde et al.\ 1999, Mundy, Looney \& Welch 2000)

B335 has been imaged with millimeter interferometers in several previous
studies (Chandler \& Sargent 1993, Hirano et al.\ 1992, Saito et al.\ 1999).
In this paper, we present observations of dust continuum emission from
B335 at both 3~mm and 1.2~mm obtained with the IRAM Plateau de Bure 
Interferometer (PdBI) that obtain considerably better sensitivity. These 
observations sample well the density profile of B335 from size scales of 
$\sim 5000$~AU to $\sim500$~AU, affording an order of magnitude better 
resolution than has been possible with single dish telescopes. The
visibilities can discriminate the behaviour of the various physical models 
that all match the near infrared extinction in the outer regions of B335.

\section{Observations} 
Continuum emission from B335 was observed with the IRAM PdBI 
simultaneously with the CS(5-4) line emission reported by Wilner et 
al.\ (2000).  In brief, a single pointing was observed using 5 antennas 
in the D configuration giving baseline lengths from the shadowing limit 
of 15 meters to nearly 80 meters. The tuning was single (lower) sideband 
at 100~GHz (3.0~mm) and double sideband at 245~GHz (1.2~mm). The half power 
field of view is $50''$ at 3.0~mm and $20''$ at 1.2~mm. The absolute flux 
scale was determined through observations of the standard source MWC349, 
assumed to be 1.09~Jy at 3.0~mm at 1.82~Jy at 1.2~mm, with estimated 
uncertainties of 20\% and 10\%, respectively. Continuum visibility records 
at the two wavelengths were formed for each 60~second integration of the
digital correlator ($3\times160$~MHz at 1.2~mm, $1\times160$~MHz at
3.0~mm), excluding a few channels contaminated by molecular line emission, 
which resulted in 2372 records at 3.0~mm and 4824 records at 1.2~mm (after 
flagging occasional bad records). In addition to amplitude and phase, each 
record also contains a variance measure, determined from the system
temperatures and antenna gains.

\section{Constructing Model Visibilities}
We analyze the dense core structure using the interferometer 
measurements directly, in the visibility domain, without producing images.  
While this approach is computationally intensive, it recognizes 
the limitations of standard Fourier inversion and deconvolution techniques,
and allows a much more direct comparison with models than analyzing images, 
avoiding the problems caused by synthesized beam characteristics.

The B335 visibilities are compared to theoretical models of protostellar  
envelope structure by constructing synthetic visibilities, taking account of 
(1) the dust continuum radiative transfer, and (2) the specifics of 
the observations, including the exact $(u,v)$ sampling and primary beam
attenuation.  The analysis of dust emission requires more complexity 
than dust extinction, because the models must include assumptions
about the temperature distribution and specific mass opacity in addition to
the model density field. Since we cannot consider an exhaustive list 
of protostellar envelope models, we instead fix attention on a few widely 
promoted density fields, including simple power law descriptions, 
inside-out collapse, and the Bonnor-Ebert sphere. These models all 
successfully match the near-infrared extinction data in the outer regions 
of B335, beyond $\sim 5000$~AU, but they predict substantial differences
in the inner regions that are sampled by the interferometer.

\subsection{Analytic Description for Power Law Distributions}
The measured visibilities are given by the Fourier transform of the 
antenna weighted intensity distribution emergent from the globule. 
In our analysis, we must numerically model visibilities for comparison 
with the data.  However, several simplifying assumptions exist that 
provide an analytic solution which affords some intuitive understanding 
of how the various physical quantities affect the visibilities. 

For optically thin dust emission, the observed intensity at impact 
parameter $b$ from a spherically symmetric globule is given by
\begin{equation}
I_{\nu}(b) = 2 \int_{b}^{R_{\mathrm{out}}} {B_{\nu}(T_d(r)) \, 
\kappa_{\nu}(r) \, \rho(r) \, \frac{r}{\sqrt{r^2 - b^2}} \, dr}
\end{equation}
where $R_{\mathrm{out}}$ is the outer radius of the globule, $T_d(r)$ the 
dust temperature, $\rho(r)$ the density, $\kappa_{\nu}$ the specific dust 
opacity, and $B_{\nu}(T)$ is the Planck function. If the density, and
temperature follow simple radial power laws, $\rho \propto r^{-p}$, 
$T \propto r^{-q}$, and the opacity is a power law in frequency that does
not vary along the line-of-sight, $\kappa_{\nu} \propto \nu^{\beta}$, then 
if the emission is assumed to be in the Rayleigh-Jeans (R-J) regime 
($h\nu/kT << 1$) from a globule with infinite outer boundary, the emergent 
intensity also has a simple power law form, 
i.e. $I(\nu,b) \propto \nu^{2+\beta} b^{-(p+q-1)}$ (Adams 1991). 
If we neglect the attenuation resulting from the antenna pattern 
of the interferometer, then the visibility distribution is also a power law, 
i.e.  $V(\nu,u) \propto \nu^{2+\beta} u^{(p+q-3)}$, where $u$ is the 
baseline length. 
With these approximations, the slope of the visibility profile --- 
like the intensity profile --- is determined solely by the sum of 
the power law indices assumed for the temperature and density. 
Note that a more steep density falloff with radius results in 
a less steep decrease in visibility with increasing baseline length. 
The relative brightness of the emission at different frequencies
is determined by $\beta$, the dust opacity index.

\subsection{Numerical Calculation of Model Visibilities}
The synthetic visibility datasets used to compare with the 
observations are produced by a series of programs written in the 
Interactive Data Language (IDL) developed by Research Systems Inc.

The first step consists of simulating the emergent intensity distribution.
Given inputs of model density and temperature distributions, and an adopted 
specific mass opacity, the program calculates the intensity image 
consisting of a $512 \times 512$ pixel array at resolution 
$0\farcs5$ per pixel, using the full Planck function for the emissivity,  
and integrates the radiative transfer equation through the model globule. 
These parameters determine the shape of the emergent intensity, and also
the relative brightness at 1.2~mm and 3.0~mm (from the frequency dependence 
of the opacity). 
In general, each model is normalized by matching to the 1.3~mm flux 
measurement of Shirley et al.\ (2000), $F=0.57\pm0.09$~Jy within 
a circular aperture (top-hat) of $40''$ diameter.
The matching is achieved by iteratively scaling the product
$\kappa_{\nu} \rho$ ($\kappa_{\nu}$ is the opacity, $\rho$ the density). 
We adopt an outer boundary of $R_{\mathrm{out}}=0.15$~pc in the models; 
this is well constrained both by the extinction measurements 
(Harvey et al. 2001) and by molecular line observations 
(e.g.\ Zhou et al.\ 1990). Modifying the outer boundary assumption
by as much as 50\% was also found to have almost no discernable effect 
on our results.

The next step is to simulate observations of the model intensity images.
The program approximates the antenna pattern of the IRAM dishes 
as a Gaussian of FWHM $50'' \: (100$~GHz$/ \nu)$. Varying illumination,
focus and pointing during the observations will cause the actual antenna
pattern of the IRAM dishes to depart slightly from a pure Gaussian, but
we expect this deviation to be small enough to have little impact on our
analysis. 
We must account for the fact that the position of the peak millimeter
emission is not exactly at the field center. To reduce the number of
parameters to be fit, we fix the offset from the center as determined 
from a simple Gaussian fit to the 1.2~mm data: $\Delta$R.A., 
$\Delta$Dec.\ $=(2\farcs23\pm0\farcs08, -0\farcs79\pm0\farcs12$),
giving coordinates for the peak of $\alpha={}$19:37:00.89, 
$\delta={}$7:34:10.0.
Experimenting with offsets $\sim 0.1''$ different from these values
showed no discernable change in the fitting results. 
The program performs an inverse FFT (to choose the positive sign in the 
exponential), and adjusts phase to take account of the offset position.
This procedure results in an array of the visibility function sampled on  
a regular grid of $u$ and $v$. The program then interpolates both the 
real and imaginary parts of the visibility to the exact $(u,v)$ points
sampled by the IRAM PdBI observations for comparison with the data.

\subsection{Model Selections}
As detailed above, the model visibilities are derived from the assumed
(1) mass density distribution $\rho(r)$, (2) dust temperature distribution 
$T_d(r)$, and (3) specific mass opacity of the dust $\kappa_{\nu}$.
We consider the expected form of these quantities, and their uncertainties.

\subsubsection{Density}
The ultimate goal is to constrain the density distribution of B335, 
given realistic choices for the temperature and opacity. The most basic
models are simple power laws with radius, with the aim to constrain the 
power law index $p$ in the inner region. As an extension we also consider
the physical models that arose in the Harvey et al.\ extinction study:
(1) a Bonnor-Ebert sphere with $\xi_{max}=12.5$, and 
(2) an inside-out collapse model with 
$R_{\mathrm{inf}}=0.03$~pc ($25''$ at 250~pc). 
We also consider the effect of the bipolar outflow on the millimeter
data, adopting the crude hollow cone description that successfully matched
the near-infrared extinction data.

\subsubsection{Temperature}
\label{sec:tempdis} 
For a optically thin dust envelope heated by a central source, 
the dust temperature distribution is expected to follow a power law 
with index determined by the frequency dependence of the opacity, 
i.e. $T_{d} (r) \propto L^{q/2} r^{-q}$, where $q=2/(4+\beta)$
(Doty \& Leung 1994). 
Breakdowns in this approximation will occur at small radii where the 
envelope becomes optically thick to the bulk of the short wavelength 
emission (or there is a circumstellar disk), 
and at large radii where heating from the Interstellar Radiation Field 
(ISRF) becomes important.

For B335, the inner breakdown of the power law approximation is expected  
at radii $\lesssim 100$~AU, and would manifest as a steeply increasing 
temperature gradient. For the IRAM PdBI observations, the longest baseline 
of $\sim 60$~k$\lambda$ corresponds to a linear resolution of $\sim 900$~AU. 
Hence, any optically thick region at small radii, or the presence of
a disk, would appear as an unresolved component in these data. 
The nature and extent of the breakdown at large radii depends on the 
local strength of the ISRF. The modelling of Shirley et al.\ (2002) suggests 
that for typical envelope density distributions $\rho \propto r^{-p}$, 
with $p \sim 1$--2, the temperature follows very closely a power law 
description until falling to just under 10~K, where it flattens until 
at large radii external heating induces a rise to $\sim 12$--15~K.
The detailed models of Shirley et al.\ (2002) for B335, using $p=1.8$ and 
an opacity index $\beta=1.8$, indicate that the temperature falls to about 
10~K at a radius of $\sim 5000$~AU. For the model PdBI visibilities, any 
(small) variations in temperature beyond this radius have little effect 
because the mass density has already decreased to 
$\sim 8 \times 10^4$~cm$^{-3}$ (only 6\% of its value at 1000~AU), and the
interferometer begins to filter emission on angular scales much larger 
than the FWHM at 1.2~mm, $\sim 20''$ ($\sim5000$~AU). 
For our standard model, we adopt the simple power law, with a minimum 
temperature of 10 K, i.e.
\begin{equation}
T_{d} (r) = 
\begin{cases}
10 \; (5000 \; \mathrm{AU}/r)^{0.4} \; \mathrm{K} \ \ & \mathrm{for \ r}  
\le 5000 \; \mathrm{AU} \\
10 \; \mathrm{K} & \mathrm{for \ r} > 5000 \; \mathrm{AU}
\end{cases}
\end{equation}
In section~\ref{sec:temperr} we investigate the uncertainties in 
the density profile that result from the temperature distribution in 
the model fitting process.

\subsubsection{Mass Opacity}
The mass opacity of dust grains in the millimeter region of the spectrum
in protostellar envelopes is uncertain, but generally assumed to follow a 
power law with frequency, $\kappa_{\nu} \propto \nu^{\beta}$.
The power law index varies depending on the dust properties, but tends to
be bounded by a small range, roughly 1--2 (Ossenkopf \& Henning 1994).
For our standard model, we follow Looney, Mundy \& Welch (2000) and 
adopt $\kappa_{\nu} = 0.1 \: (\nu / 1200$~GHz$)$~cm$^2$~g$^{-1}$, a power 
law with dust emissivity index $\beta$ equal to unity. Since the globule is
optically thin at the wavelengths of the PdBI observations, especially
at 3.0~mm, the power law index in the models can be adjusted to produce 
the best simultaneous match at the two observed wavelengths.

\section{Method for Fitting Model Parameters and Evaluating Fit Quality}
The basic procedure is to maximize the probability distribution:
\begin{equation}
 P(\mbox{Model} \: | \: \mbox{data})=
\prod_i{e^{-(Z_i-f(x_i; p, m))^2/2\sigma_i^2}}
\times e^{-(m - m_0)^2/2\sigma_m^2}
\end{equation}
where the $Z_i$ are the visibility data points with uncertainty $\sigma_i$,
$f(x_i; p, m)$ are the model data points, $p$ a free parameter in the
models, and $m$ a model parameter about which we have a constraint (namely
that it is a Gaussian random variable with mean $m_0$ and standard deviation
$\sigma_m$). Maximizing the probability distribution is equivalent to
minimizing the logarithm of its inverse. Taking account the fact that the
$Z_i$ are by nature complex visibilities, we want to minimize a modified
$\chi^2$:
\begin{equation}
\tilde{\chi}^2 = \sum_i{\frac{|Z_i-f(u,v; p,m)|^2}{\sigma_i^2}} + 
\frac{(m-m_0)^2}{\sigma_m^2}
\end{equation}
It is useful to be more explicit about the parameters $p$ \& $m$. 
The free parameter $p$ is used to describe the shape of the model, 
e.g.\ the index of the power-law density distribution. 
The parameter $m$ allows us to include observational uncertainties,
e.g.\ the 20\% uncertainty in the normalization of the model derived
by matching the 1.3~mm flux measured by Shirley et al.\ (2000).
This is achieved by allowing the 1.2~mm model visibilties to be scaled 
by a constrained parameter $m$, a Gaussian random variable with mean 
1.0 and standard deviation 20\%. For the 3.0~mm model visibilities, we 
use two different approaches: (1) scale the 3.0~mm data by the same 
factor $m$ as the 1.2~mm data to force a dust opacity index of unity;
(2) allow the 3.0~mm data to be freely scaled, to measure the dust opacity 
index and as a general check for consistency (by interpreting the derived 
value $\beta$ in the context of the realistic range described above).

We select a model description and perform a fit for the model parameters  
by minimizing the modified $\chi^2$ distribution which includes 
data at both wavelengths. Since the models are non-linear in the fitting
parameters, we analyze the uncertainty in the best-fit model parameters
using a Monte Carlo technique known as the {\em bootstrap} method 
(Press et al.\ 1992, p.\ 691). To describe this method, consider the 
dataset $S$ used in the fitting. The first step is to construct a new 
dataset of equal size $S^{\prime}$, by randomly selecting $N$ times from  
the original dataset.  This new sample is then analyzed in the same way 
as the original dataset, and the fitted model parameters recorded. 
This process is then repeated $n$ times, where $n$ is sufficiently large
that the resulting distribution of best-fit model parameters is
insensitive to its exact value (for our models, typically $n=200$).
If the fitting parameters may be successfully constrained by the data, then 
the distribution of best-fit parameters is very close to Gaussian, and the
standard deviation of the distribution provides an estimate for the  
uncertainty of the parameters that best fit the original dataset. 

The bootstrap method proves especially powerful for this analysis
because small variations in the value of the modified $\chi^2$ as defined 
above do not necessarily represent small variations in fit quality. 
The visibility dataset comprises a very large number of very low
signal-to-noise measurements, which has the effect that the $\chi^2$ wells 
are extremely shallow, and two models may seem almost equally good, 
despite the fact that there is ample signal to distinguish them. 
This occurs because the individual visibilities do not distinguish between 
the two models (represented in the $\chi^2$ well being shallow) but the 
combination of all the individual visibilities does (represented by a lower 
$\chi^2$ corresponding to a better fit).
This issue could be circumvented by binning the 
visibilities e.g.\ radially, to increase 
the signal-to-noise, and then performing a $\chi^2$ fit to the binned 
values. We avoid this solution because the binning loses information on the
asymmetry 
of the globule that is hidden in the dataset by the vectorial nature of 
visibility data. For the spherically symmetric models, we bin the data
for graphical comparison with the models.

The shallowness of the $\chi^2$ wells results in the minimum values of 
the reduced $\chi^2$ for various models to all have similar absolute
values, in the range $\chi^2_{\nu}=1.4$--$1.5$.
While a lower value of $\chi_{\nu}^2$ may be interpreted as providing 
a better fit, the small range of variation confuses the differences in 
the fit quality. Therefore, we adopt graphical techniques to 
express the superiority of one model fit over another.

\section{Results and Analysis}
Table~1 summarizes the results from the $\chi^2$ fits to several
spherically symmetric models, as well as axisymmetric models.
For fits marked `\#a', the normalization of the 3.0~mm visibilities 
is constrained by fixing the dust opacity index $\beta$ to unity. 
For fits marked `\#b', the dust opacity index is included as an 
additional fitting parameter.

\subsection{Spherically Symmetric Models (Fits I--IV)}
Although the B335 dense core shows clear departures from spherical
symmetry when viewed both in dust extinction and molecular emission, 
for simplicity and insight we start by investigating spherically 
symmetric density models. 
In Fits~Ia \& Ib, we use a power law model for the density distribution.  
In Fit~IIa, we use a power law density model but include an
additional point source component to the intensity distribution. We also
consider Bonnor-Ebert spheres (Fit~IIIa; Bonnor 1956, Ebert 1955), and the 
Shu (1977) inside-out collapse model (Fit~IVa).

\subsubsection{Single Power Laws}
In the simplest model, the density distribution is described by a 
single power law. 
The best-fitting model has a power law index of $p=1.65 \pm 0.05$ 
(Fit~Ib) with a dust opacity index $\beta=0.81 \pm 0.12$ that is consistent
with unity. Fit~Ia, which assumes $\beta=1$, gives a density power law
index that agrees within uncertainties with that for Fit~Ib. 
Figure~1 shows a plot of binned visibility amplitude vs.\
 $(u,v)$ distance for both wavelengths. The binning is logarithmic and 
oversamples the data by a factor of 2; filled symbols and non-filled symbols
are not completely independent. The amplitudes are the vectorial 
mean of the complex visibilities in each bin (corrected for the offset 
phase center), the error bars represent one standard deviation in the mean.
For the dust emission at both wavelengths, the decrease in visibility
amplitude with baseline length may be well described by a power law. 
The plots also show the visibility curves for the best fit model 
($p=1.65$) and models that deviate by $\pm2\sigma$ ($p=1.55$ and $p=1.75$). 
These models have not been scaled by the fitting parameters since this 
term contributes to the $\chi^2$ and its exclusion allows the differing 
quality of each fit to be better seen (note that the $\pm2\sigma$ models  
require a scaling at 1.2~mm of 1.19 and 0.75 for $p=1.55$ and $p=1.75$,
respectively).

\subsubsection{Sensitivity to Temperature Assumptions}
\label{sec:temperr}
These results use the temperature distribution described in 
section~\ref{sec:tempdis}, which is based on a physical argument. 
To understand the sensitivity of the density power law results to
the temperature, we investigate how subtle quantitative changes 
affect the results by repeating the fitting procedure with temperature 
distributions that (1) have different (constant) temperature in the
outer envelope, and (2) have different power law indices.

We find that the fitted density power law index is insensitive to 
(small) changes in the envelope temperature: variations of $\pm 2$~K lead 
to variations in the inferred density slope of much less than the quoted  
uncertainty. The main reason is that the antenna FWHM is only $20''$
at 1.2~mm (which has the higher signal-to-noise ratio of the two 
wavelengths), 
and the data are not very sensitive to the temperature structure at 
larger radii. Therefore, uncertainties in the strength of the ISRF are 
relatively unimportant in this study. 

Changes in the power law index of the inner temperature distribution 
produce more significant changes in the fitting results. The globule 
envelope is optically thin at both wavelengths; if the R-J approximation
were also to hold, then the inferred density power law index $p$ would be 
related to the temperature power law index $q$ by the relation 
$p+q=$~constant. However, the R-J approximation breaks down, especially at 
1.2~mm, with the consequence that $p$ is more sensitive to changes in 
$q$. We find that varying $q$ by $\pm 0.1$ leads to variations in $p$ at the 
level of $\simeq \mp 0.15$.

\subsubsection{Sensitivity to an Unresolved Central Component}
\label{sec:pointerr}
As already noted, there may exist a central region of B335 for which
the envelope will be optically thick at infrared wavelengths, causing the
temperature profile to 
depart from a simple power law. In this region, the temperature is expected
to climb very sharply with radius, resulting in increased emission at long 
baselines beyond the extrapolation of a simple power law. The presence of 
an unresolved disk, effectively a warm central point source, would 
have a similar effect.  While the PdBI data samples insufficient 
baseline lengths to directly measure the flux of any such compact component, 
it will, if present, flatten the visibility profiles of the various models 
and therefore affect the fitted density power law index. 
Fit IIa quantifies the effect using the combination of the simple power law 
for density, temperature and mass opacity from Fit Ia, but including an 
additional point source component with $F \propto \nu^3$ (i.e.\ the spectrum
of a source where the majority of the emission is optically thin at 1.2~mm,
with a dust opacity index of unity).
For a point source flux $F=12\pm7$~mJy at 1.2~mm, the most allowed by   
the measurements at long baselines, the fitted density index is reduced 
to $p=1.47 \pm 0.07$. This point source flux corresponds to an implied disk
mass of $M \simeq 2 \times 10^{-3} \: \mathrm{M}_{\odot} \: 
(60$~K$/T_{\mathrm{disk}})$ for our adopted opacity law. This is at the lower
end of the 0.002--0.3~M${}_{\odot}$ range of disk masses that have been 
observed around T Tauri stars by Beckwith et al.\ 1990. Since accretion disks
result from conservation of angular momentum during the collapse process,
the small inferred mass of the disk may be related to the fact that B335 has
a relatively low rotation rate (Frerking, Langer \& Wilson 1987). However, 
accretion disks may also be connected with the driving source of bipolar 
outflows, and the B335 outflow is not unusual, with size, momentum and
energy typical for a low mass protostar (Goldsmith et al.\ 1984).

When a compact source of emission is included in the analysis, the fitted
density power law index is reduced by $\sim -0.2$. Thus, possible point 
source ``contamination'' is the dominant source of systematic uncertainty
in the fitted density structure. This is also the case in single dish
measurements, where a point source can add a systematic uncertainty of up to
$\sim -0.5$ to the inferred density index (e.g.\ Shirley et al.\ 2002).

Table~\ref{tab:errors} summarizes various sources of systematic error 
and the level of uncertainty that the variations contribute to the 
fitted density power law index in the spherically symmetric models.

\subsubsection{Comparison With Previous Studies}
The best fit density power law index of $p=1.65$ is slightly less steep 
than the value of $p=1.8$ found by Shirley et al.\ (2002) from SCUBA data.
It is perhaps significant that the results from the two instruments are 
consistent to within the uncertainties considering the resolution of the
JCMT at 850~$\mu$m is $15''$, roughly encompassing the entire range of 
radii probed by the the IRAM PdBI. However, even taking account all sources 
of uncertainty, the best-fit power law index of $p=1.65$ is significantly
less steep than the $p=2.0$ power law index of the hydrostatic 
isothermal sphere that so successfully matches the outer envelope 
of B335 in near-infrared extinction (Harvey et al.\ 2001) 
and dust emission shape (Shirley et al.\ 2002)
This result suggests that the density profile becomes less steep in the 
inner regions ($r \lesssim 25''$), a conclusion also suggested by the 
extinction data, where two models with this character (inside-out collapse 
and the Bonnor-Ebert sphere) provided better fits than a single power law.

\subsubsection{Broken Power Laws, Inside-Out Collapse, Bonner-Ebert Sphere}
We have also investigated models with density distributions that flatten  
in the inner region:
(1) a broken power law, with fixed index of $p=2$ in the envelope, 
with model parameters the break radius and inner power law index; 
(2) an inside-out collapse model with model parameter the infall radius 
(this is approximately a special case of the broken power law);
and (3) a Bonnor-Ebert sphere, where the model parameter is the
dimensionless outer radius of the sphere $\xi_{\mathrm{max}}$.
None of these models fit the data nearly as well as a single power law. 

For the broken power law model, it proved impossible to obtain robust 
results for the two fitted parameters, which are coupled and have 
comparable effects in the simulated datasets. Decreasing the break radius, 
or increasing the power law index in the inner region; each steepens 
the apparent density profile, and it is not possible to distinguish 
between the two effects. 

The inside-out collapse model and the Bonnor-Ebert sphere that 
matched the extinction data both provide poor fits to the 
visibility data that sample the envelope closer to the protostar. 
Figure~2 shows a comparison between the binned visibility amplitude 
vs. $(u,v)$ distance and the curves produced from these models:
the inside-out collapse model with $R_{\mathrm{inf}}=0.03$~pc, 
and the Bonnor-Ebert sphere with $\xi_{\mathrm{max}}=12.5$. 
There are large discrepancies between these models and the data. 
Because these models so poorly describe the data, 
the $\chi^2$ statistic provides a poor discriminant of model parameters.  
At the $\chi^2$ distribution minimum, the bootstrap calculation bounces 
between the end points of the grid at each iteration, no matter 
how wide a range of model parameters are considered. 
Essentially, the data prefer a model with a single power law index 
$p \simeq 1.65$, and neither the inside-out collapse model nor 
the Bonner-Ebert sphere model can produce an equivalent slope over 
a sufficiently large enough range in spatial scales. 

The inclusion of a point source component does not rectify the 
problems in fitting with these two physical models, despite effectively 
reducing the slope of the fitted density distribution.
The Bonner-Ebert sphere has a relatively flat density distribution 
at small radii, and therefore the visibilities drop sharply at
long baselines. Adding a point source to the Bonnor-Ebert sphere model
adds intensity at long baselines but still cannot produce a good fit
over the large range in spatial scale sampled by the data. 
For the inside-out collapse model, the fit fails to constrain the 
infall radius because this parameter does not affect the slope of the 
density distribution in the region where the data provide constraints
(e.g.\ $R_{\mathrm{inf}} \gtrsim 0.024$~pc $=20''$ at 250~pc). 

A combination of the inside-out collapse model with a point source 
also produces a poorer fit to the data than does a simple power law 
with a point source. The $\chi^2$ of this model is equivalent to 
power laws that are nearly $3 \sigma$ away from the best fit 
(i.e. the fit quality is equivalent to the power law models with
$p=1.30$ or $p=1.65$, when the best fit is for $p=1.47 \pm 0.07$).  
This seems a surprising result, since the standard impression
of the inside-out collapse model is that it behaves like a power law with
index very close to 1.5 in the infalling region. But, while this is 
asymptotically true at very small radii, the local power law index of the 
inside-out collapse solution in fact decreases as one moves outwards in the 
infalling region, attaining a minimal value close to unity just within the 
infall radius. This means that for the same total flux within a $40''$
diameter, the inside-out model has greater intensity at larger radii than
does a simple $p=1.5$ power law. The effect is that while the slopes of the 
two visibility profiles are similar in the spatial range where the PdBI data
are sensitive, the primary beam attenuation (FWHM $\simeq 20''$ at 1.2~mm)
causes the inside-out model visibilities to be 25\% fainter than the single
power law model. The 
required increase in the scaling parameter, $m$, to best match the data is 
significant and contributes to the increased value of $\chi^2$, resulting in 
a worse fit for the inside-out collapse model. Essentially, the inside-out 
collapse model cannot simultaneously match the Shirley et al.\ (2000) flux 
measurement and the normalization of our observed visibility profile.

\subsection{Axisymmetric Models (Fit V)}

In Figure~3, left column of panels show images of B335 made from 
the 1.2~mm and 3~mm visibilities, respectively. These images 
show that the B335 dense core is not spherically symmetric at this
size scale, but rather is more like an oblate spheroid, flattened 
along the axis of the bipolar outflow (nearly East-West on the sky). 
In Figure~3, the second column of panels shows images of the best fit 
spherically symmetric models, made by processing the synthetic 
visibilities in the same way as the data. The aspect ratio of the 
emission is clearly different between the models and the data.
The asymmetry may be caused by the action of the outflow, or it may
reflect a flattening in the intial configuration of the core.

For the dust extinction data at scales $\gtrsim 3500$~AU, the asymmetry 
was well reproduced by a model that assumed that the outflow 
had hollowed out a bipolar cone, with best fit constant semi-opening 
angle $\alpha=41^{\circ} \pm 4^{\circ}$. To extend the analysis of 
the visibility data, we have added the same simple representation of 
the bipolar outflow to the power law density distribution models to 
determine if this model can also successfully explain the asymmetry in 
the observed dust emission. In these models, we assume no 
unresolved point source of significant flux, having already quantified 
the effect of this additional component in section~\ref{sec:pointerr}.
We also assume that the temperature remains a simple radial power law.
This is an extreme simplification as the emptied outflow cavities provide
a less obscured view to the hotter inner region, and as a result
the temperature of dust near to the boundary of the outflow cavity
will likely have been underestimated.

Fit~V (a \& b) repeats the recipe of the $\chi^2$ analysis, using a 
density model that comprises a radial power law with index $p$ and 
a hollow-cone outflow with constant semi-opening angle $\alpha$. 
As with Fit~Ia, Fit~Va assumes a dust opacity index of unity, whereas 
Fit~Vb includes the dust opacity index as an additional parameter.
As listed in Table~1, the two fits give nearly identical results, 
both with each other and with the outcome of Fit~I. 
The best fit power law index (Fit~Vb) is $p=1.65 \pm 0.05$, with 
outflow semi-opening angle $\alpha=40^{\circ} \pm 5^{\circ}$,
and dust opacity index $\beta=0.74 \pm 0.10$. 

In Figure~2, the additional columns of panels show images at both 
wavelengths made from the synthetic visibilities of the best fit model, 
and those with parameter values that differ by $2 \sigma$ from the 
best fit,
i.e. $p=1.55$, $\alpha=30^{\circ}$ and $p=1.75$, $\alpha=50^{\circ}$. 
The best fit spherically symmetric model is also shown to demonstrate 
the effect of adding the outflow cones. The quality of the match to 
the data is visible in (1) the extent of the apparent asymmetry, 
and (2) the number and spacing of the intensity contours. 
Note that the density gradient parameterized by $p$ and the degree of 
asymmetry parameterized by $\alpha$ are nearly independent.  
The outflow model with $p=1.65$ and 
$\alpha=40^{\circ}$ provides the best match to the imaged data, and 
has lower reduced $\chi^2$ despite the additional fitting parameter.

The best fit density power law index is essentially unchanged from the 
spherically symmetric analysis. That these results should be so similar
is partly a consequence of the adopted outflow geometry.
The main effect of the outflow cone is to change the normalization of the 
intensity profile along the direction through the cone, but not to change 
the shape or power law index of the profile. This would not be the case for 
an outflow model without a constant opening angle. 
For instance, for a model where the opening angle decreases with distance 
from the protostar (e.g.\ a parabola), the intensity profile along the 
outflow axis becomes less steep as relatively less material is removed 
from the core at larger radii. The fitted power law index in such a model 
would be higher than for the simple cone model. As an example, consider
an outflow geometry where the semi-opening angle is $50^{\circ}$ at 
500~AU, decreasing to $30^{\circ}$ at 5000~AU; in this model, the intensity 
profile along the outflow axis has power law index that is $\sim 0.15$ less 
steep than for the constant opening angle case, or the profile along an axis 
perpendicular to the flow. While we have not analyzed this class of model
in detail, since the outflow covers less than two quadrants of the sky,
spatial averaging of the regions affected by the outflow and the regions 
that are unaffected suggests an uncertainty of roughly 
$\delta p \lesssim 0.15/2 \sim 0.08$, a small effect on fitted power laws.

The good agreement in the fitted opening angles for the simple cone model
in this work and the extinction study of Harvey et al.\ (2001) may not be
especially significant.  While the consistency is encouraging, it most 
likely reflects the fact that the crude outflow model creates a detectable 
degree of oblate flattening only for semi-opening angles near $45^{\circ}$.
For much largers opening angles, too much of the dense core is removed 
by the outflow and the core becomes disk like; for much smaller angles, 
the effect on the core is small, except in a small region at the poles.

\subsection{Discussion}

The analysis presented in the previous sections provides strong evidence
that the density distribution power law index is less than 2.0 in the 
central $5000$~AU ($20''$) region where the PdBI observations are sensitive. 
The best fit single power law for the density distribution has index of 
$p=1.65 \pm 0.05$. This value should be considered an upper limit since
including a central point source --- the dominant contributor of systematic 
uncertainty --- can reduce the value to $p=1.47 \pm 0.07$. 
The remaining sources of systematic uncertainty considered, of which 
the most important is the radial dependence of the temperature distribution, 
likely contribute a total uncertainty at the level $\delta p \lesssim 0.2$.

The Harvey et al.\ (2001) extinction study demonstrated a density index
of very close to 2.0 ($p=1.91 \pm 0.07$) over a $3500$--$25000$~AU range
in radius, indicative of an isothermal hydrostatic envelope, a result
consistent with several different analyses of dust emission from single dish
observations that probe similar spatial scales (Shirley et al.\ 2002, 
Motte \& Andr\'{e} 2001).
The new interferometric dust emission data, which probes smaller scales,
provides strong evidence that the density profile becomes less steep 
in the inner $5000$~AU of B335. This conclusion agrees with the hints 
from detailed modelling of dust extinction indicative of a turnover in 
the density profile at a radius of about $6500$~AU, as well as application 
of the Shu (1977) inside-out collapse model to molecular line observations 
(Zhou et al.\ 1993, Choi et al.\ 1995).

The inferred density profile of the B335 core conforms closely to 
the predictions in standard theories of isolated star formation 
(Shu 1977, Larson 1969, Penston, 1969) where an envelope with 
density power law index of $p=2.0$ surrounds an inner region
with a $p=1.5$ distribution characteristic of material freely falling
under the force of gravity. 
However, even with the contribution of a unresolved point source 
component, the specific Shu (1977) inside-out collapse model provides 
a worse fit to the data than does a simple power law. The local density index 
within the turnover (infall) radius in the inside-out collapse model is 
significantly less steep than $p=1.5$, ranging from roughly unity to $1.5$. 
Consequently, the inside-out collapse model cannot successfully 
match both the normalization of the visibility profile in the 
IRAM PdBI observations and the total flux within a large aperture 
measured with a single dish by Shirley et al.\ (2000). 
The constraint provided by the single dish measurement --- effectively the  
visibility information at zero-spacing that cannot be measured directly
by the interferometer --- is crucial to this analysis; without it, 
the density distribution power law index can be constrained only to within 
$\delta p \sim \pm 0.3$, not including contributions from systematic errors.

The images of B335 presented in Figure~3 demonstrate clear departures from 
spherical symmetry. The inner part of the globule appears significantly 
extended perpendicular to the outflow axis. This is in contrast 
to images of dust emission with lower angular resolution that show nearly 
spherical symmetry. While there are no robust predictions for the effect
of the outflow on the density distribution, combining the power law density 
model with a very simple bipolar cone to describe the outflow improves
the match to the data, but does not alter significantly the inferred 
shape of the radial variation in density. 

Observations of B335 at longer baselines are needed to constrain the flux 
from any point source component that may be present at the center of the  
dense core. Such a component might be expected in the form of a circumstellar
disk. Indeed, the maximal 1.2~mm flux of an unresolved component 
($\sim 12$~mJy) lies on the low end of the range expected for the B335 disk 
based on the subarcsecond submillimeter fluxes for Class 0 sources observed
with the CSO-JCMT interferometer (Brown et al.\ 2000), perhaps related to 
the relatively low rotation rate of B335 (see Zhou 1995).

In the future, high resolution observations of B335 at submillimeter 
wavelengths may help to mitigate the uncertainties associated with the
temperature distribution. Spatial information on the optically thin emission 
distribution at a third wavelength that is sufficiently short to be outside 
the Rayleigh-Jeans regime for the majority of the dense core should allow 
for partly breaking the degeneracy between temperature, density and opacity.
Such observations may be obtained with the Submillimeter Array (SMA), now
under construction, or the Atacama Large Millimeter Array (ALMA).

\subsection{A Physical Density Distribution for B335}

Throughout the analysis presented so far, attention has been paid only to 
the power law index of the density distribution; the normalization has been 
absorbed in satisfying the constraint placed by the single dish 1.3~mm 
flux measurement of Shirley et al.\ (2000). To allow the results to be used 
in future modeling of B335, we now discuss the physical scale of the 
density distribution. 
The present study, in concert with the Harvey et al.\ (2001) extinction 
work, affords the unique position to propose a density distribution that 
satisfies both dust emission and extinction data over a range in radius from 
500~AU to 25,000~AU.

For the adopted temperature distribution and mass opacity at 1.2~mm, 
the density within $r \lesssim 20''$ (5000~AU) is:
\begin{equation}
n_H(r) \simeq
\begin{cases}
8.1 \times 10^{5} \: \mathrm{cm}^{-3} \left( \frac{r}{10^3 \: \mathrm{AU}} 
\right)^{-1.65} & \mathrm{for \ no \ point \ source }\\
7.6 \times 10^{5} \: \mathrm{cm}^{-3} \left( \frac{r}{10^3 \: \mathrm{AU}} 
\right)^{-1.47} & \mathrm{for \ 12 \ mJy \ point \ source}
\end{cases}
\end{equation}
where $n$ represents the total number density in the gas 
($n=n(H_2) + n(He) +{}$\ldots${}=\rho/(\mu m_H)$, $\mu=2.29$).
Note that including the outflow as a hollow cone with semi-opening angle 
$40^{\circ}$ increases by 14\% the coefficient of the density distribution 
power law (and leaves the index unchanged).

For the envelope beyond $r \gtrsim 26''$ (6500~AU), the Harvey et al.\ 
(2001) extinction work suggested an isothermal, hydrostatic, structure with 
$r^{-2}$, consistent with the results of the dust emission studies 
of Shirley et al.\ (2002) and Motte \& Andr\'{e} (2001). This envelope 
density distribution can be adopted into our model without affecting the 
inferred structure in the inner regions, since emission from the envelope
is a negligible contribution at small impact parameters ($\lesssim 20''$), 
and is filtered out at large impact parameters by the antenna pattern 
($20''$ FWHM at 1.2~mm).
However, the correct normalization of the density distribution in the outer 
envelope relative to the inner regions is difficult to assess because 
the uncertainties involved in the extinction work (gas-to-dust ratio, 
reddening law) are different from those that arise in the dust emission 
studies (specific mass opacity, temperature structure), and 
the previous dust emission studies do not resolve the inner regions.

The analysis of the interferometer dust continuum data allows us to draw 
some conclusions about the behaviour of the density profile in the 
transition region between the two regimes, and hence to infer a consistent
scale for the envelope.
In order for any model to simultaneously match the normalization of the
observed visibility profile and the Shirley et al.\ (2000) measurement 
of the flux within $r < 20''$, the local power law index in the transition
region cannot decrease from the value measured in the inner region. 
The power law index must remain constant or become steeper.
A simple continuation of the inner power law until reaching the envelope 
implies a density distribution for $r > 26''$ (6500~AU):
\begin{equation}
n(r) \simeq 
\begin{cases}
1.6 \times 10^{4} \: \mathrm{cm}^{-3} \: 
\left( \frac{r}{10^4 \: \mathrm{AU}} \right)^{-2.0} &
\mathrm{for \ no \ point \ source} \\
2.2 \times 10^{4} \: \mathrm{cm}^{-3} 
\left( \frac{r}{10^4 \: \mathrm{AU}} \right)^{-2.0} &
\mathrm{for \ 12 \ mJy \ point \ source}
\end{cases}
\end{equation}
For the case of no point source, this composite density distribution 
gives an average power law index of $p=1.90$ over the 3500--25000~AU 
radius range sampled by the dust extinction data, precisely reproducing 
the extinction result ($p=1.91 \pm 0.07$), and it gives an average 
power law index of $p=1.80$ over the 1000--25000~AU radius range, 
which reproduces the result of Shirley et al.\ (2002). Remarkably, the 
normalization of this envelope distribution is also within 6\% of the
singular isothermal sphere.

\section{Summary}

We present a study of the density distribution of the protostellar collapse 
candidate B335 using continuum observations at 1.2~mm and 3.0~mm made 
with the IRAM PdBI. In summary:

\begin{enumerate}
\item We perform a detailed analysis of the interferometer measurements
directly in the visibility domain, not in the image domain. Though 
computationally intensive, this approach avoids the limitations of 
standard Fourier inversion and deconvolution algorithms.

\item The PdBI visibility data strongly constrain the density distribution
at radii from $\sim 500$~AU to $\sim 5000$~AU from the protostar. 
Within this inner region, the density distribution is well described by 
a single power law.  The best fit power law index $p=1.65 \pm 0.05$ 
(1~$\sigma$)
is significantly less steep than values close to $p=2.0$ derived for larger 
radii from analyses of dust emission and extinction. We consider the effects 
of various sources of systematic uncertainty on the derived value of $p$.
Including a central source of point-like emission in the model,
as might come from an unresolved accretion disk, reduces the power law index 
to $p=1.47 \pm 0.07$. The remaining sources of systematic uncertainty, of
which the most important is the radial dependence of the temperature
distribution, likely contribute a total uncertainty at the level of 
$\delta p \lesssim 0.2$.

\item The inferred density profile for B335, with a power law index
close to $r^{-1.5}$ near the protostar within an $r^{-2}$ envelope, 
matches the generic paradigm of gravitational collapse for isolated, 
low mass star formation. 
However, the specific inside-out collapse solution of Shu (1977) does 
not produce as good a fit to the dust emission data in the inner region
as does a simple single power law, largely because of the 
shallow slope ($p \sim 1$) of the inside-out collapse solution 
just within the infall radius.

\item Images made from the visibility data show clear departures from 
spherical symmetry, with the inner part of the B335 core elongated 
perpendicular to the outflow axis. Including the outflow in the model
fitting as a hollowed bipolar cone improves the match to the data and
does not significantly change the derived radial density power law index  
from the value obtained in the spherically symmetric analysis. 
An outflow geometry where the opening angle decreases with distance from  
the protostar location would result in a slightly larger power law index, 
roughly $\delta p \lesssim 0.05$.

\item Observations that probe the subarcsecond millimeter continuum 
structure of B335 are needed to constrain the central point source 
contribution to the dust emission.

\end{enumerate}

\acknowledgements
DWAH thanks Jonathan Jenkins for many useful discussions concerning
statistics.
We acknowledge the IRAM staff from the Plateau de Bure and from Grenoble
for carrying out the observations and for their help during the data reduction.
We are especially grateful to Roberto Neri for his assistance.
Partial support for this work was provided by NASA Origins of Solar Systems
Program Grant NAG5-6266.

\clearpage
% The tables and figures begin here....
\begin{deluxetable}{lllll}
\tablenum{1}
\tabletypesize{\footnotesize}
\tablewidth{0pt}
\tablecolumns{5}
\tablecaption{Summary of the Density Model Fits \label{tab:fits}}
\tablehead{ \colhead{Fit} & \colhead{Density Model} 
& \colhead{Dust Index} & \colhead{Fitted Model Parameter(s)} & 
\colhead{$\chi^2_\nu$}}
\startdata
Ia & Power law: $\rho \propto r^{-p}$ & $\beta=1$ & $p=1.61 \pm 0.06$; 
$M=1.10 \pm 0.13$ & 1.4647 \\
Ib & As Fit Ia & $\beta=$Free & $p=1.65 \pm 0.05$; 
$\beta=0.81 \pm 0.12$ & 1.4646 \\
IIa & Power law $+$ point flux $F$ & $\beta=1$ & $p=1.47 \pm 0.07$;
$F=12 \pm 7$~mJy; $M=1.24 \pm 0.15$ & 1.4645\\
IIIa & Bonner-Ebert, $\xi_{\mathrm{max}}=12.5$ & $\beta=1$ & \ldots &
1.5025 \\
IVa & Inside-out, $R_{\mathrm{inf}}=0.03$~pc & $\beta=1$ & \ldots & 1.4674\\
Va & Power law $+$ outflow & $\beta=1$ & $p=1.60 \pm 0.06$; 
$\alpha=38^{\circ} \pm 5^{\circ}$; $M=1.10 \pm 0.13$ & 1.4625 \\
Vb & As Fit Va & $\beta=$Free & $p=1.65 \pm 0.05$; 
$\alpha=40^{\circ} \pm 5^{\circ}$; $\beta=0.74 \pm 0.10$ & 1.4622 \\
\enddata
%\tablecomments{See text for explanation}
\end{deluxetable}

\clearpage
\begin{deluxetable}{lll}
\tablenum{2}
\tabletypesize{\footnotesize}
\tablewidth{0pt}
\tablecolumns{3}
\tablecaption{Summary of systematic uncertainties in the fitted density
power-law index $p$ \label{tab:errors}}
\tablehead{ \colhead{Model Parameter} & \colhead{Variation} 
& \colhead{Resulting $\delta p$ Systematic Error}}
%\rotate
\startdata
Temperature power-law index, $q$ & $\delta q \lesssim \pm 0.1$ & 
$\lesssim \mp 0.15$\\
Outer envelope temperature (ISRF) & $\delta T \lesssim \pm 2$~K & 
$\lesssim \pm 0.05$\\
Central point source, flux $F$ & $F = 12 \pm 7$~mJy & 
$-0.18 \mp 0.1$\\
Dust opacity spectral index $\beta$ & $\beta=1 \rightarrow \beta={}$free & 
$\simeq 0.05$\\
Outer boundary of B335, $R_{\mathrm{out}}$ & 
$\delta R_{\mathrm{out}} / R_{\mathrm{out}} \lesssim 50$\% &
$\lesssim 0.01$\\
Outflow cone geometry, $\alpha=\alpha (r)$ &
$\delta \alpha (r) \lesssim \pm 20^{\circ}$\tablenotemark{1} &
$\lesssim \mp 0.08$ 
\enddata
\tablenotetext{1}{Refers to an outflow geometry where the opening angle
changes monotonically by $20^{\circ}$ over the $500$--$5000$~AU range in
radius to which the data is sensitive.}
\end{deluxetable}

\clearpage
%Figures

\begin{figure}[htb]
\figurenum{1}
\setlength{\unitlength}{1in}
\begin{picture}(6,6.0)
\put(-0.2,-1.5){\includegraphics{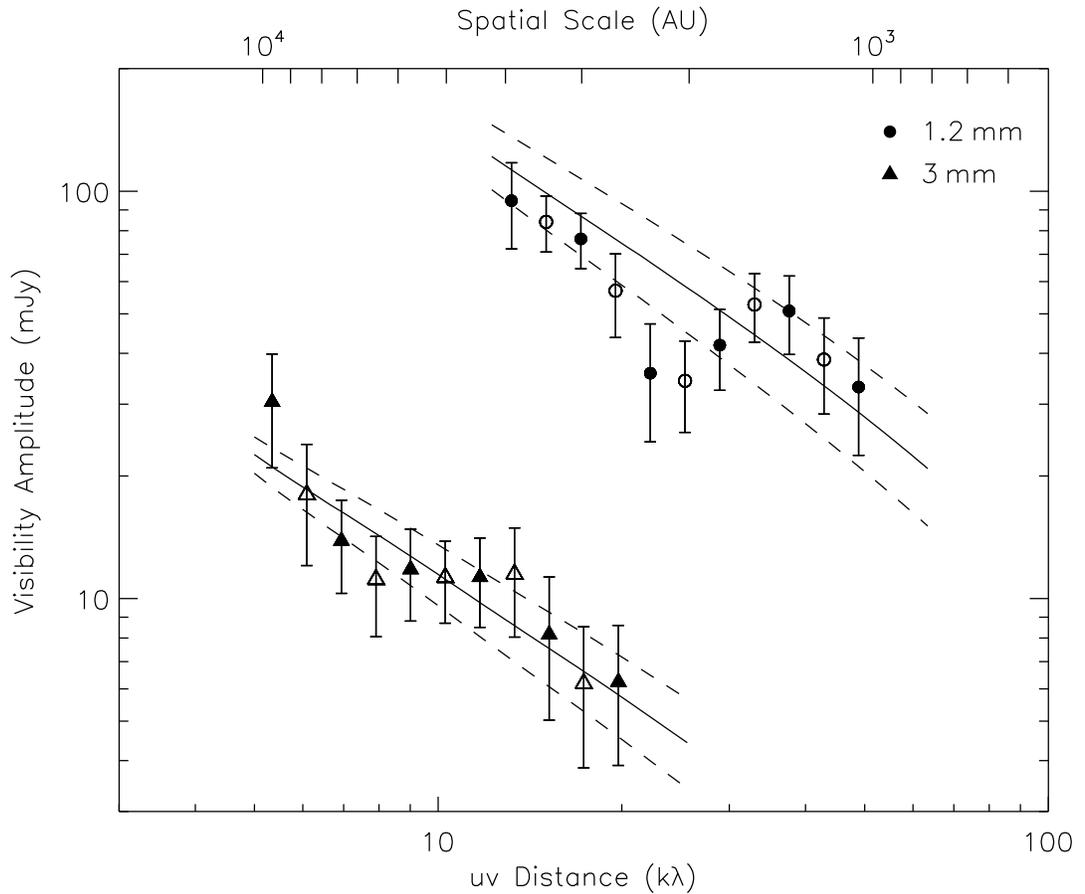}}
\end{picture}
\caption{Binned visibility amplitudes against $(u,v)$ distance at 3.0~mm
(triangles) and 1.2~mm (circles) for the PdBI observations of B335
together with the unscaled curves for the best fitting single power law 
model $p=1.65$ (solid line), and power law models that deviate by 
2~$\sigma$,  $p=1.55$ (lower dashed line) and $p=1.75$ (upper dashed line).
Note that the bins oversample the data and the filled symbols are not 
completely independent from non-filled symbols. The spatial scale on the top 
is defined as 250~pc${} \times \lambda / D$, where $D$ is the baseline 
length.}
\end{figure}

\clearpage

\begin{figure}[htb]
\figurenum{2}
\setlength{\unitlength}{1in}
\begin{picture}(6,6.0)
\put(-0.2,-1.5){\includegraphics{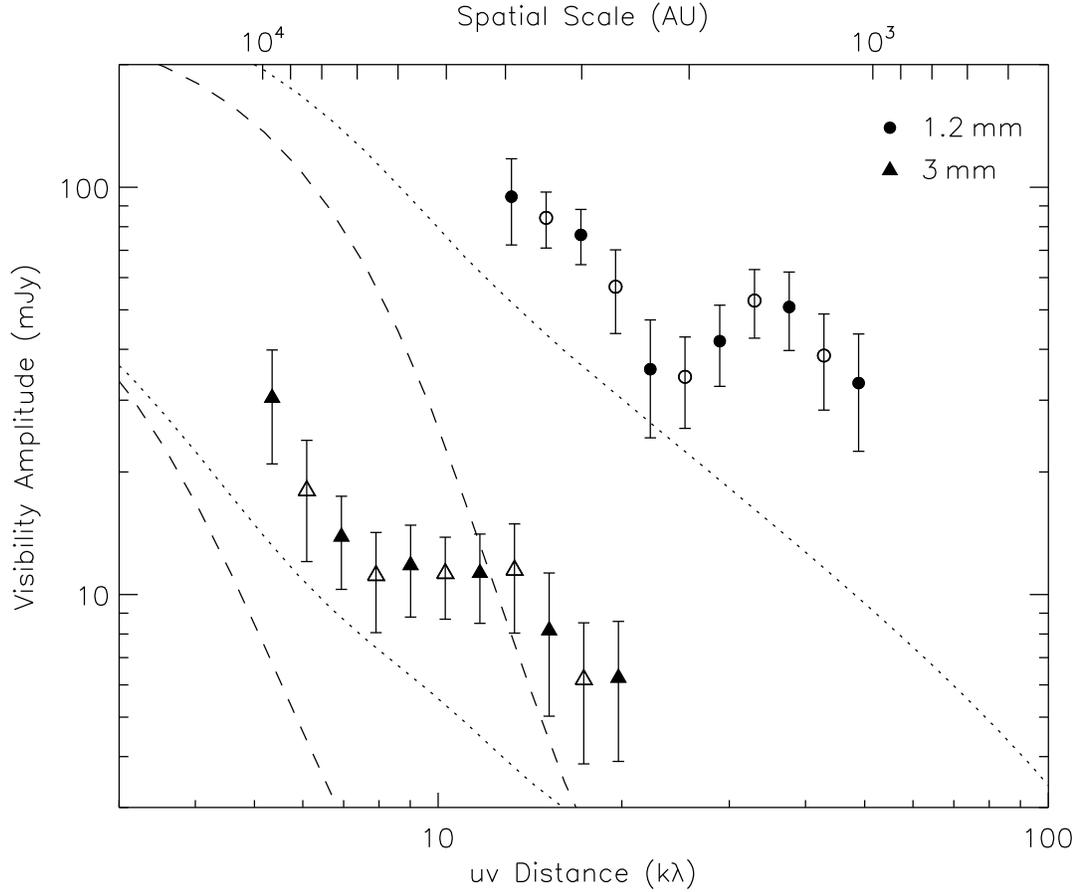}}
\end{picture}
\caption{Binned visibility amplitudes against $(u,v)$ distance at 3.0~mm
(triangles) and 1.2~mm (circles) for the PdBI observations of B335
together with theoretical curves for the Bonnor-Ebert sphere with 
$\xi_{\mathrm{max}}=12.5$ (dashed line) and the inside-out collapse model 
with $R_{\mathrm{inf}}=0.03$~pc (dotted line). Neither of these models can 
successfully match the data. The bins oversample the data 
and the filled symbols are not completely independent from non-filled
symbols. The spatial scale on the top is defined as 
250~pc${} \times \lambda / D$, where $D$ is the baseline length.}

\end{figure}

\clearpage

\begin{figure}[htb]
\figurenum{3}
\setlength{\unitlength}{1in}
\begin{picture}(6,6.5)
\put(-0.2,-0.7){\includegraphics{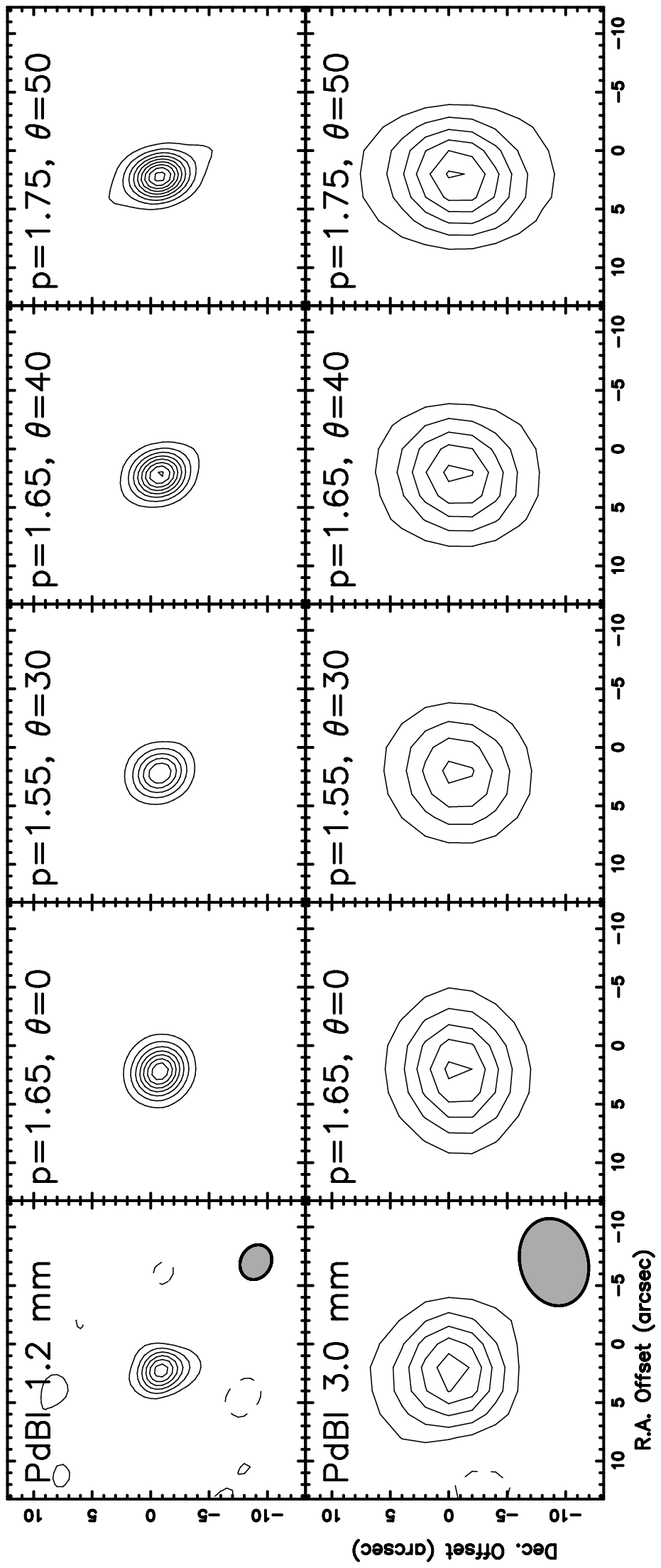}}
\end{picture}
\caption{Images of B335 at 1.2~mm (top panels) and at 3.0~mm (lower 
panels). The left column shows images made from the PdBI data, and
the columns to the right show images made from synthetic datasets 
produced from the various protostellar envelope models,
including the best fit spherically symmetric power law model, $p=1.65$,
and the best-fit asymmetric model, $p=1.65$, $\alpha=40^{\circ}$,
bracketed by asymmetric models with parameter values that differ by
$2\sigma$, i.e. 
$p=1.55$ and $\alpha=30^{\circ}$, $p=1.75$ and $\alpha=50^{\circ}$. 
The contour levels are $(2,4,6,8,...)\times$ the rms noise of the 
imaged data at each wavelength, 3.0~mJy for the 1.2~mm panels
and 0.8~mJy for the 3.0~mm panels.
The synthesized beam sizes are shown in the lower right corners of
the imaged data in the first column.}
\end{figure}


\clearpage
\begin{references}

\reference{}
Adams, F.C.\ 1991, \apj, 382, 544

\reference{} 
Beckwith, S.V.W., Sargent, A.J., Chini, R.S.\ \& G\"{u}sten, R.\ 
1990, \aj, 99, 924

\reference{}
Bonnor, W.\ 1956, \mnras, 116, 351

\reference{}
Brown, D.W., Chandler, C.J., Carlstrom, J.E., 
Hills, R.E., Lay, O.P., Matthews, B.C., Richer, J.S.\ \&  
Wilson, C.D.\ 2000, \mnras, 319, 154

\reference{}
Caselli, P., Walmsley, C.M., Zucconi, A., Tafalla, M., Dore, L.\ \&
Myers, P.C.\ 2002, \apj, 565, 331

\reference{}
Chandler, C.J.\ \& Sargent, A.L.\ 1993, \apj, 414, 29

\reference{}
Choi, M., Evans, N.J.II, Gregerson, E.M., Wang, Y.\ 
1995, \apj, 448, 742

\reference{}
Doty, S.D.\ \& Leung, C.M.\ 1994, \apj, 424, 729

\reference{}
Ebert, R.\ 1955, {\em Z.Astrophys.}, 37, 217

\reference{}
Frerking, M.A., Langer, W.D.\ \& Wilson, R.W.\ 1987, \apj, 313, 320

\reference{}
Goldsmith, P.F., Snell, R.L., Hemeon-Heyer, M.\ \& Langer, W.D.\ 1984,
\apj, 286, 599

\reference{}
Harvey, D.W.A., Wilner, D.J., Lada, C.J., Myers, P.C., Alves, J.F.\
\& Chen, H.\ 2001, \apj, 563, 903

\reference{} 
Hirano, N., Kameya, O., Kasuga, T.\ \& Umemoto, T.\ 1992,
\apj, 390, 85

\reference{} 
Hogerheijde, M.R., van Dishoeck, E.F., Salverda, J.M.\ \& Blake, G.A.\
1999, \apj, 513, 350

\reference{}
Keene, J., Davidson, J.A., Harper, D.A., Hildebrand, R.H., Jaffe, 
D.T., Loewenstein, R.F., Low, F.J.\ \& Pernic, R.\ 1983, \apj, 
274, 43

\reference{}
Keene, J.\ \& Masson, C.R.\ 1990, \apj, 355, 635

\reference{}
Larson, R.B.\ 1969, \mnras, 145, 271

\reference{}
Looney, L.W., Mundy, L.G.\ \& Welch W.J.\ 2000, \apj, 529, 477

\reference{}
Motte, F.\ \&  Andr\'{e}, P.\ 2001, \aap, 365, 440

\reference{}
Mundy, L.G., Looney, L.W.\ \& Welch W.J.\ 2000, in
{\em Protostars and Planets IV}, eds.\ V.\ Mannings, A.P.\ Boss and
S.S.\ Russell, (Tucson: University of Arizona Press), p.\ 355

\reference{}
Ossenkopf, V.\ \& Henning, T.\ 1994, \aap, 291, 943

\reference{}
Penston, M.V.\ 1969, \mnras, 144, 425

\reference{}
Press, W.H., Teukolsky, S.A., Vetterling, W.T.\ \& Flannery, B.P.\
1992, {\em Numerical Recipies in C}, p.\ 691

\reference{}
Saito, M., Sunada, K., Kawabe, R., Kitamura, Y.\ \& Hirano, N.\
1999, \apj, 518, 334

\reference{}
Shirley, Y.L., Evans, N.J.II\ \& Rawlings, J.M.C.\ 2002, 
{\em Astro-Ph, 0204024}

\reference{}
Shirley, Y.L., Evans, N.J.II, Rawlings, J.M.C.\ \& Gregerson, 
E.M.\ 2000, \apjs, 131, 249

\reference{} 
Shu, F.H. 1977, \apj, 214, 488

\reference{}
Tafalla, M., Myers, P.C., Caselli, P., Walmsley, C.M.\ \& Comito, C.\
2002, \apj, 569, 815

\reference{}
Tomita, Y., Saito, T.\ \& Ohtani, H.\ 1979, \pasj, 31, 407

\reference{}
Visser, A.E., Richer, J.S.\ \& Chandler, C.J.\ 2001, \mnras, 323,
257

\reference{}
Ward-Thompson, D., Scott, P.F., Hills, R.E.\ \& Andre, P.\ 1994,
\mnras, 268, 276

\reference{}
Ward-Thompson, D., Motte, F.\ \& Andr\'{e}, P.\ 1999, \mnras,
305, 143

\reference{}
Wilner, D.J., Myers, P.C., Mardones, D.\ \& Tafalla, M.\ 2000, 
\apj, 544, 69 

\reference{} 
Zhou, S., Evans, N.J.II, Butner, H.M., Kutner, M.L.,
Leung, C.M.\ \& Mundy, L.G.\ 1990, \apj, 363, 168

\reference{} 
Zhou, S., Evans, N.J.II, Kompe, C.\ \& Walmsley, C.M.\ 1993, 
\apj, 404, 232

\reference{} 
Zhou, S. 1995, \apj, 442, 685

\end{references}
\end{document}